# 3D Axial-Attention for Lung Nodule Classification

## Mundher Al-Shabi[a,1], Kelvin Shak[a], Maxine Tan[a,c]

a) Electrical and Computer Systems Engineering Discipline, School of Engineering, Monash University Malaysia, 47500 Bandar Sunway, Selangor, Malaysia
b) Department of Biomedical Imaging, University of Malaya, 50603 Kuala Lumpur, Malaysia
c) School of Electrical and Computer Engineering, University of Oklahoma, Norman, OK 73019, USA

## Abstract

**Purpose**: In recent years, Non-Local based methods have been successfully applied to lung nodule classification. However, these methods offer 2D attention or limited 3D attention to low-resolution feature maps. Moreover, they still depend on a convenient local filter such as convolution as full 3D attention is expensive to compute and requires a big dataset, which might not be available.

**Methods**: We propose to use 3D Axial-Attention, which requires a fraction of the computing power of a regular Non-Local network (i.e., self-attention). Unlike a regular Non-Local network, the 3D Axial-Attention network applies the attention operation to each axis separately. Additionally, we solve the invariant position problem of the Non-Local network by proposing to add 3D positional encoding to shared embeddings.

**Results**: We validated the proposed method on 442 benign nodules and 406 malignant nodules, extracted from the public LIDC-IDRI dataset by following a rigorous experimental setup using only nodules annotated by at least three radiologists. Our results show that the 3D Axial-Attention model achieves state-of-the-art performance on all evaluation metrics, including AUC and Accuracy.

**Conclusions**: The proposed model provides full 3D attention, whereby every element (i.e., pixel) in the 3D volume space attends to every other element in the nodule effectively. Thus, the 3D Axial-Attention network can be used in all layers without the need for local filters. The experimental results show the importance of full 3D attention for classifying lung nodules.

## 1. Introduction

Lung cancer is the leading cause of cancer death in 2020, whereby it is alone responsible for 25% of all cancer deaths [1]. Examining patients during early stages using computed tomography (CT) scans can reduce lung cancer mortality [2]. The CT examination is usually performed by radiologists, whereby they locate and identify lung nodules. However, this process is time-consuming and tedious because of the large number of nodules that need to be examined.

Many deep learning-based Computer-Aided Diagnosis (CAD) systems have been introduced for lung nodule classification [3–6]. Al-Shabi et al. [6] proposed multiple paths each with different dilated convolutions and a subnetwork to guide the features through the paths. Shen et al. [5] proposed including perturbations in the training dataset whilst training deep learning models. Ren et al. [4] proposed a manifold regularized classification deep neural network (MRC-DNN) and classified nodules based on their manifold representation. Onishi et al. [7] used generative adversarial networks (GAN) to generate multiplanar images of the nodules; then, they used the generated images to train a convolutional neural network (CNN) model.

---

[1] Corresponding author, email: mundher.al-shabi @ monash.edu

However, lung nodule size variability is one of the biggest challenges, whereby the nodule diameter can be anywhere between 3 to 30 mm [6]. Recently, Non-Local based methods were proposed for scale-invariant classification of the lung nodule [8, 9]. Nevertheless, the Local-Global network [8] requires converting the nodule's three-dimensional (3D) volume to 2D patches. Such conversion leads to losing some vital information about the shape of the nodule. Moreover, these Non-Local based methods [8, 9] are not purely attention-based, and still require local filters such as CNN. To overcome these limitations, we propose 3D Axial-Attention, which is a full/ complete 3D attention model. In contrast with computationally expensive Non-Local based models, the 3D Axial-Attention is lightweight and can be applied at all layers without the need for local filters. Overall, our contributions to this work can be summarized as follows:

1. We generalize the 2D Axial-Attention to 3D and apply it for lung nodule classification.
2. We propose to use shared embeddings and 3D positional encodings.
3. Our proposed method outperforms state-of-the-art methods on the LIDC-IDRI dataset.

## 2. Methods

### 2.1. Non-Local networks

Given an input feature map $X \in \mathbb{R}^{C \times Z \times W \times H}$, with number of channels C, depth Z, height H, and width W, the Non-Local operation [10, 11] (i.e., self-attention) is defined as follows:

$$v(X) = W^v X \tag{1}$$

$$q(X) = W^q X \tag{2}$$

$$k(X) = W^k X \tag{3}$$

$$B = softmax(q(X)^\top k(X)) \tag{4}$$

$$\Psi = v(X) B \tag{5}$$

where $v(X), q(X), k(X) \in \mathbb{R}^{D \times N}$ are linear embeddings of $X$, and $W^v, W^q, W^k \in \mathbb{R}^{D \times C}$ are learnable parameters that can be implemented efficiently using 3D convolution with a kernel size of $1 \times 1 \times 1$. $N = Z \times W \times H$, and D is the size of the embedding. That is, we convert the embeddings into a 2D matrix first; then, we apply the Non-Local operation.

Equation 4 generates the attention maps of size $N \times N$. This means that every N in the output $\Psi$ is aggregated from N values in $v(X)$. Hence, it is called a Non-Local network. As compared to a local filter like CNN, which has a small receptive field (i.e., $3 \times 3$ or $5 \times 5$), the Non-Local network has a receptive field of size $Z \times W \times H$. The $q(X)^\top k(X)$ multiplication is very expensive to compute and requires $O(N^2)$ space and computational time to generate the attention map of $q(X)^\top k(X)$ and another $O(N^2)$ to apply the attention map B to $v(X)$. Notice we ignore D for simplicity as D is much smaller than N (see Table 1). Considering our data is 3D, namely $N = Z \times W \times H$, the overall computation overhead is a quadradic $O(Z^2 \times W^2 \times H^2)$. For a simple nodule with a size of $32 \times 32 \times 32$, the computation overhead is 1,073,741,824. With this significant overhead, in the literature, this limits Non-Local to be applied to one layer with low resolution (e.g., $8 \times 8 \times 8$) in 3D Dual-Path [9], 2D nodule patches as in Local-Global network [8], or by treating the depth as a channel in ProCAN [12].

### 2.2. 3D Axial-Attention

Recently, (Ho et al., 2019) [13] proposed Axial-Attention to reduce the computation overhead of the Non-Local network, as shown in Figure 1. In Axial-Attention, instead of converting the input into 2D after the embeddings $\mathbb{R}^{D \times N}$, we keep them in 4D dimensions $\mathbb{R}^{D \times Z \times W \times H}$. Then, we apply the Non-Local network normally as in Equation 4. This results in an Attention map of size $H \times H$ called Heightwise-Attention, which means every element attends to its column. In comparison, the Non-Local network generates a $N \times N$ attention map, and every element attends to every other element. Next, we apply the Non-Local operation one more time, but this time we reshape the input into $\mathbb{R}^{D \times Z \times H \times W}$. In simple words, we just bring the W-axis in front, called Widthwise-Attention. By applying Equations 4-5 to the W-axis, every element attends to its row.

In this paper, we generalize the Axial-Attention operation from 2D into 3D by adding Depthwise-Attention. That is, after Widthwise-Attention, we reshape the output matrix into $\mathbb{R}^{D\times H\times W\times Z}$ and then pass it to Equations 4-5, as depicted in Figure 2. As a result of the three operations (Heightwise, Widthwise, and Depthwise-Attention), every element in the 3D dimension space attends indirectly to every other element. The overall computation overhead of the 3D Axial Attention is $O(H \times H + W \times W + Z \times Z)$ in computation time and space. Considering $H = W = Z = \sqrt[3]{N}$, the computation complexity becomes $O(N\sqrt{N})$ which is exponentially much smaller than $O(N^2)$. For the previous example, with a nodule of size $32 \times 32 \times 32$, the computational complexity in 3D Axial-Attention is just 5,931,641 compared to 1,073,741,824 in $O(N^2)$. These huge savings in computational power (>99.4%) make the 3D Axial-Attention more applicable to all layers instead of only a few layers. This is the first attempt to apply 3D attention to all layers, whereby every element in the 3D space can attend to any other element. Compared with Local-Global networks [8], whereby we could use only two 2D Non-Local layers, the 3D Axial-Attention method implements 3D attention in all six layers.

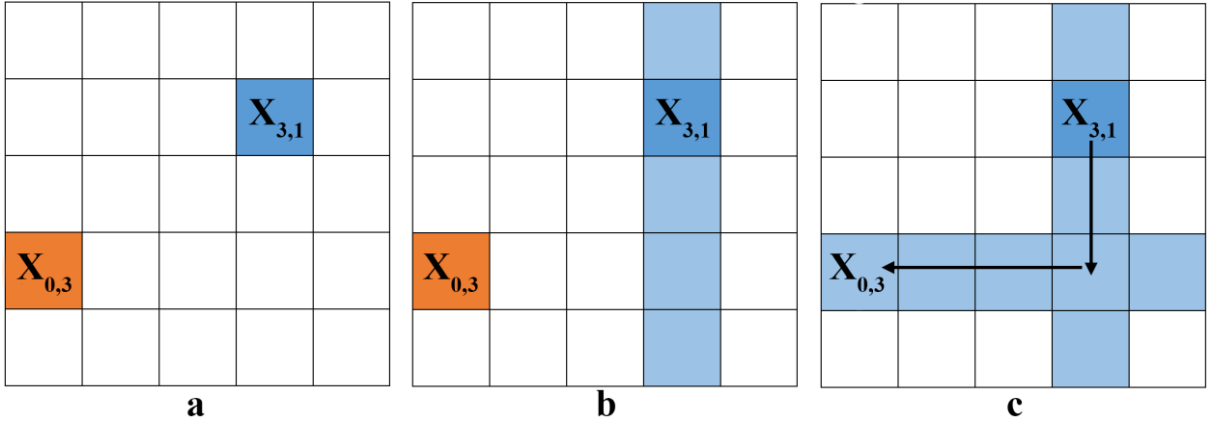

Figure 1: Consider $X_{3,1}$ wants to attend to $X_{0,3}$ (a), first we apply Heightwise-Attention (b) where every element attends its column. Here, we are just highlighting the fourth column. Next, we apply Widthwise-Attention (c), where every element attends to its row. Here, we are just highlighting the fourth row. Notice by applying the attention twice, every element can attend to any other element in 2D space.

## 2.3. Shared Embeddings & Positional Encodings

Another issue with Non-Local networks is the redundancy between the three embedding functions $v(X), q(X), k(X)$. This problem becomes crucial when we train our network with a small dataset such as LIDC-IDRI [14]. The overparameterization in generating the embedding features leads to overfitting. Therefore, we propose to use shared embeddings – We replace the three embedding functions with just one. We rewrite Equations 4-5 with shared embeddings:

$$\boldsymbol{\Psi} = h(X)softmax(h(X)^{\intercal}h(X)) \qquad (6)$$

$$h(X) = q(X) + \boldsymbol{P} \qquad (7)$$

where $h(X)$ is the sum of the shared embeddings $q(X)$ and the positional encoding $\boldsymbol{P}$. $\boldsymbol{P} \in \mathbb{R}^{D\times Z\times W\times H}$ is a 3D positional encoding that is added to the embeddings to explicitly encode the position. The position encoding is crucial as the Non-Local operation is position-invariant [11]. In this work, we propose $\boldsymbol{p}_{i,j,u} \in \boldsymbol{P}$, a 3D positional encoding which is made up from three learnable vectors $\boldsymbol{r}^z, \boldsymbol{r}^h, \boldsymbol{r}^w \in \mathbb{R}^D$ whereby they represent the depth, height, and width positions, respectively:

$$\boldsymbol{p}_{i,j,u} = r_i^z + r_j^h + r_u^w \qquad (8)$$

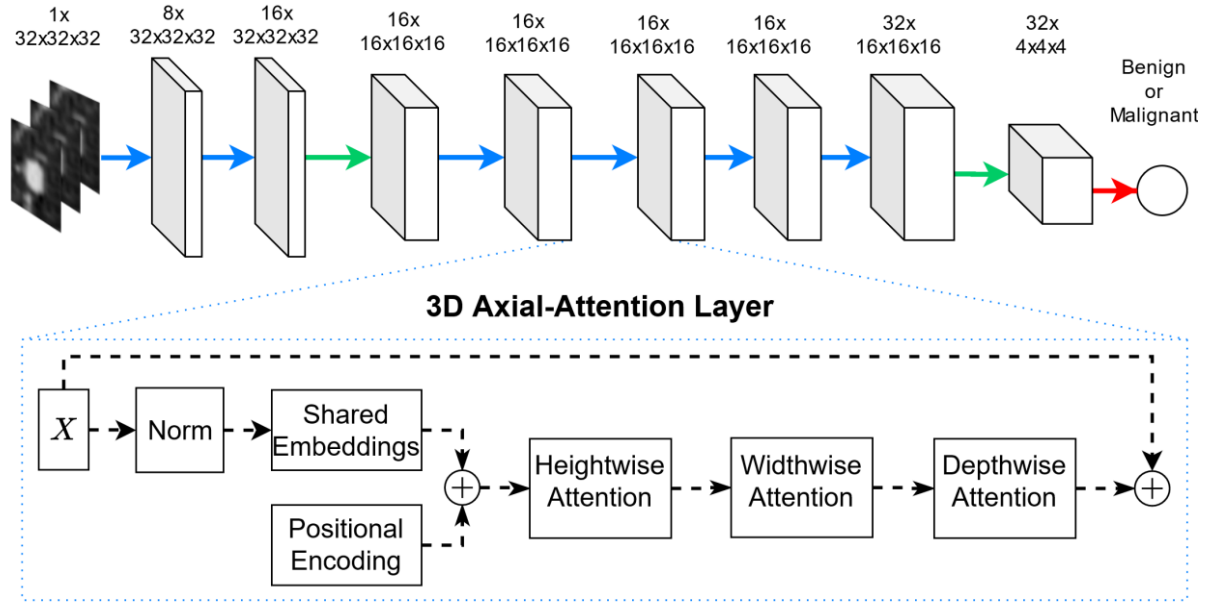

*Figure 2: Illustration of the proposed network architecture and new 3D Axial-Attention network. X denotes the input of the layer, and Norm represents Layer Normalization. Identity indicates a simple connector.*

Before the embedding function, we normalize the input using Layer Normalization [15], as shown in Figure 2. Unlike Batch Normalization [16], Layer Normalization is not sensitive to the batch size, which allows us to increase the network's depth at the expense of reducing the batch size without decreasing performance.

### 2.4. Network Architecture & Training Details

Table 1 tabulates the details of the whole network architecture. We used six 3D Axial-Attention layers with different D sizes, whereby D controls the number of channels. In between the 3D Axial-Attention layers, we used 3D Max-Pooling to reduce the dimensionality as we increase the number of channels. Finally, we flattened the multidimensional input to 1D before passing it to a Fully Connected layer with a Sigmoid activation function. We used dropout [16] with a probability of 0.5 just before the Fully Connected layer to prevent overfitting.

We trained the network using Adam optimizer [17] with a Binary Cross-Entropy loss function and a batch size of 64. We used a weight decay of 0.0001 for the Fully Connected layer. Next, we trained the network for 20 epochs with a learning rate of 0.001, followed by an additional 40 epochs with a learning rate of 0.0001. All the learnable parameters were initialized using Xavier uniform initialization [18].

*Table 1: Details of the 3D Axial-Attention network architecture. Kr denotes Kernel Size, and St denotes Stride Size*

| Layer | Input dimensions ($C \times Z \times W \times H$) | Output dimensions | Notes |
| --- | --- | --- | --- |
| 3D Axial-Attention: 1 | 1×32×32×32 | 8×32×32×32 | D = 8 |
| 3D Axial-Attention: 2 | 8×32×32×32 | 16×32×32×32 | D = 16 |
| 3D Max-Pooling | 16×32×32×32 | 16×16×16×16 | Kr. = (2,2,2), St. = 2 |
| 3D Axial-Attention: 3 | 16×16×16×16 | 16×16×16×16 | D = 16 |
| 3D Axial-Attention: 4 | 16×16×16×16 | 16×16×16×16 | D = 16 |
| 3D Axial-Attention: 5 | 16×16×16×16 | 16×16×16×16 | D = 16 |
| 3D Axial-Attention: 6 | 16×16×16×16 | 32×16×16×16 | D = 32 |
| 3D Max-Pooling | 32×16×16×16 | 32×4×4×4 | Kr. = (4,4,4), St. = 4 |
| Fully Connected | 32×8×8×8 | 1 | Dropout = 0.5 |

### 2.5. Dataset

We used the LIDC-IDRI dataset [14] consisting of 1,018 CT scans, and up to four radiologists annotated each scan. We only utilized the scans annotated by at least three radiologists [6, 8, 12]. Cases with a median rating of more than three are considered malignant, while cases with a median rating of less than three are considered benign. We discarded the cases which have median ratings of exactly three. This results in 848 nodules, of which 442 were benign and 406 were malignant.

Each nodule is normalized to have an isotropic resolution in all three dimensions. Hence, every pixel in the 3D volume has a 1 mm$^3$ resolution. Then we cropped a 32×32×32 (32 mm$^3$) volume around each nodule. For data augmentation, we rotated each nodule at four different angles (0, 90, 180, and 270 degrees) around the three axes (x, y, and z). Then we split the data into 10-fold cross-validation subsets. Finally, we applied standard scalar normalization.

## 3. Results

Comparisons of our proposed method against state-of-the-art lung nodule classification methods are summarized and tabulated in Table 2. Additionally, we compare 2D and 3D Axial-Attention. We observe that our proposed 3D Axial-Attention method outperforms other state-of-the-art methods on all evaluation metrics. Furthermore, it surpasses other Non-Local based methods including 3D DPN [9], Local-Global [8], and 2D Axial-Attention.

In Figure 3, we show samples of randomly selected images of nodules, whereby the first and the second row show correctly-classified nodules, and the third row shows the misclassified nodules.

Table 2: Performance comparisons of our proposed 3D Axial-Attention method with state-of-the-art methods on the LIDC-IDRI dataset

| Model | AUC | Accuracy | Precision | Sensitivity |
|---|---|---|---|---|
| HSCNN [19] | 85.6 | 84.2 | --- | 70.5 |
| Multi-Crop [3] | 93.0 | 87.14 | --- | 77.0 |
| Local-Global [8] | 95.62 | 88.46 | 87.38 | 88.66 |
| Gated-Dilated [6] | 95.14 | 92.57 | 91.85 | 92.21 |
| 3D DPN(Ensemble)[9] | --- | 90.24 | --- | 92.04 |
| MRC-DNN [4] | --- | 90.0 | --- | 81.0 |
| Perturbated DNN [5] | 91.0 | 83.0 | --- | 90.0 |
| 2D Axial-Attention | 94.74 | 91.63 | 92.19 | 90.15 |
| 3D Axial-Attention | **96.17** | **92.81** | **92.59** | **92.36** |

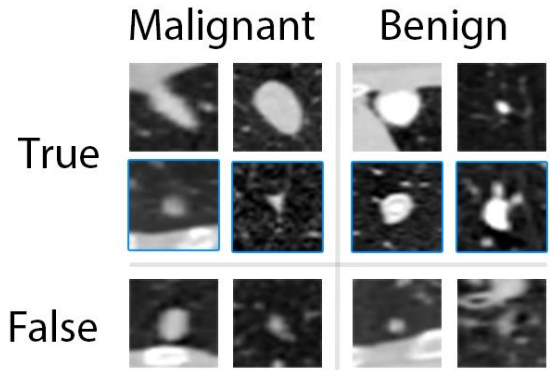

Figure 3: Randomly selected images of nodules with their ground truth labels (True/ False) and the predicted labels (Malignant/ Benign) from our proposed model. The first two rows depict the correctly-classified nodules, whereas the last row depicts the misclassified nodules. Blue borders donate images of mid-range nodules that are correctly classified by 3D Axial-Attention and misclassified by the Local-Global network.

## 4. Conclusion

In this study, we propose a new 3D Axial-Attention method for lung nodule classification. The proposed method provides full computationally efficient 3D attention. Moreover, we propose shared embeddings and 3D positional encodings. Our proposed method's lightweight design allows us to apply it to all layers without any local filters. The results show that our proposed method outperforms all state-of-the-art methods on the LIDC-IDRI dataset.

## Acknowledgments

This work was supported by the Fundamental Research Grant Scheme (FRGS), Ministry of Education Malaysia (MOE), under grant FRGS/1/2018/ICT02/MUSM/03/1, the Electrical and Computer Systems Engineering and Advanced Engineering Platform, School of Engineering, Monash University Malaysia and the TWAS-COMSTECH Joint Research Grant, UNESCO.